\documentclass{sigchi}

\usepackage{balance}       
\usepackage{graphics}      
\usepackage[T1]{fontenc}   
\usepackage{txfonts}
\usepackage{mathptmx}
\usepackage[pdflang={en-US},pdftex]{hyperref}
\usepackage{color}
\usepackage{booktabs}
\usepackage{textcomp}
\usepackage{dirtytalk}
\usepackage{csquotes}
\usepackage{subfig}

\usepackage{microtype}        
\usepackage{ccicons}          

\usepackage{todonotes}

\def\plaintitle{Lessons Learned from Developing a Microservice Based Mobile Location-Based Crowdsourcing Platform}
\def\plainauthor{}

\def\plainkeywords{Authors' choice; of terms; separated; by
  semicolons; include commas, within terms only; this section is required.}

\makeatletter
\def\url@leostyle{%
  \@ifundefined{selectfont}{
    \def\UrlFont{\sf}
  }{
    \def\UrlFont{\small\bf\ttfamily}
  }}
\makeatother
\def\pprw{8.27in}
\def\pprh{11.69in}

\definecolor{linkColor}{RGB}{6,125,233}
\hypersetup{%
  pdftitle={\plaintitle},
pdfauthor={\plainauthor},
  pdfkeywords={\plainkeywords},
  pdfdisplaydoctitle=true, 
  bookmarksnumbered,
  pdfstartview={FitH},
  colorlinks,
  citecolor=black,
  filecolor=black,
  linkcolor=black,
  urlcolor=linkColor,
  breaklinks=true,
  hypertexnames=false
}

\begin{document}

\title{\plaintitle}

\numberofauthors{4}
\author{
  \alignauthor{Irwyn Sadien\\
    \affaddr{Xi'an Jiaotong-Liverpool University}\\
    \affaddr{Suzhou, China}\\
    \email{irwyn.sadien@xjtlu.edu.cn}}\\
  \alignauthor{Konstantinos Papangelis\\
    \affaddr{Xi'an Jiaotong-Liverpool University}\\
    \affaddr{Suzhou, China}\\
    \email{k.papangelis@xjtlu.edu.cn}}\\
      \alignauthor{Charles Fleming\\
    \affaddr{University of Mississippi}\\
    \affaddr{Mississippi, USA}\\
    \email{fleming@cs.ucla.edu}}\\
      \alignauthor{Hai-Ning Liang\\
    \affaddr{Xi'an Jiaotong-Liverpool University}\\
    \affaddr{Suzhou, China}\\
    \email{HaiNing.Liang@xjtlu.edu.cn}}\\
}

\maketitle

\begin{abstract}
Research in Mobile Location-Based Crowdsourcing is hindered by a marked lack of real-world data. The development of a standardized, lightweight, easily deployable, modular, composable, and most of all, scalable experimentation framework would go a long way in facilitating such research. Conveniently, these are all salient characteristics of systems developed using a microservices approach. We propose QRowdsource - a MLBC experimentation framework built using a distributed services architecture. In this paper, we discuss the design and development of QRowdsource, from the decomposition of functional components to the orchestration of services within the framework. We also take a look at how the advantages and disadvantages of using a microservices approach translate to our specific use case and deliberate over a number of lessons learned while developing the experimentation framework.  
\end{abstract}

\section{Introduction and Background}
Crowdsourcing, a term first coined in 2006, refers to \say{a distributed problem-solving model in which a crowd of undefined size is engaged to solve a complex problem through an open call} \cite{6216342}. This has evolved from being a few targeted implementations (such as Amazon Mechanical Turk and Wikipedia) to a practice that engages millions of people worldwide through numerous platforms and applications. Crowdsourcing is considered as a problem-solving tool~\cite{chiu_what_2014}, an online distributed problem-solving and production model~\cite{brabham_crowdsourcing_2009,Corsar2017,Corsar-citizen-sensing}, an open collaborative learning paradigm~\cite{unsw_australia_business_school_organizational_2014,a_campo_community_2019}, and a new resource for product development~\cite{poetz_value_2012, khan_macrotask_2019, macro-def}. As network technologies and computing systems have become more and more embedded in physical and social contexts, crowdsourcing has taken on new forms that leverage the ubiquity of mobile devices to engage crowds 'on-the-go' in order to collect and validate huge amounts of data with spatio-temporal characteristics. While there has been a lot of research on crowdsourcing in the conventional sense, much less attention has been paid to mobile location-based crowdsourcing (MLBC), and this is partly due to the fact that there are very few real-world implementations of MLBC the provide open access to their data. As Zhao and Han state:
 
\par
\begin{displayquote}\textelp{} in reality, there is no available data set that can be utilized directly in [MLBC] tasks. Thus, from an engineering perspective, it is a necessity to collect real-world data to help validate different research ideas and systems instead of running simulations over synthetic data \cite{Zhao2016SpatialDirections}. 
\end{displayquote}

In order to pave the way for future research in this paradigm of crowdsourcing, it is thus important to ensure that a proper experimental framework be developed and made available. The challenge lies in developing lightweight, easily deployable, robust, reliable, composable, and most importantly scalable systems to support MLBC. The trend today in achieving such goals is to move away from conventional monolithic architectures towards a distributed set of loosely coupled modules, known as microservices.
 
A microservice is commonly defined as a unit of code built around a singular business functionality that runs in its own environment and communicates through a standard interface \cite{AbdollahiVayghan2018DeployingLearned}. The advantages of using microservices over a monolithic architecture are numerous: 1) Microservices are independent, small in size and implement a limited number of functionalities. This makes them easy to test and debug in isolation from the rest of the system. 2) Changing any part of a microservice architecture does not require that the whole system be taken down. Redeployment, upgrades, and maintenance leads to very short redeployment times. This also leads to a high degree of composability in implementing different functional systems using existing modules. 3) Scaling a microservice based application does not require scaling of the whole system, but rather only those components experiencing heavy load. What is more, with the elasticity of cloud computing, these services can be adaptively scaled to make the most efficient use of resources. 4) The only constraint imposed on the interoperability of microservices is a common interface for communication. This means that developers are free to choose the most optimal combination of technologies in implementing microservices \cite{Dragoni2017Microservices:Tomorrow}.

In this paper, we discuss the design and development of a microservice based experimentation platform for MLBC, titled QRowdsource. We provide a description of the QRowdsource platform and identify the functional process in MLBC in order to break them down into services, discussing the data considerations and intercommunication mechanisms required, and giving an overview of the software stack used. Finally, we make a case for microservice architecture in developing an experimentation framework for MLBC by looking at the specific advantages and disadvantages, and talk about the lessons learned in this application.

\section{QRowdsource}
We developed QRowdsource; a MLBC platform that leverages QR-codes as location markers and entry points into the crowdsourcing ecosystem. The QRowdsource project itself is an investigation into Proximity Technology (PT) enabled MLBC ``in-the-wild'', with the first research experiment for which the platform was configured at the time of writing being a study on the suitability of different types of tasks for PT-enabled MLBC.

\subsection{QRowdsource Workflow} 
The front-facing component of QRowdsource was developed as a Progressive Web Application (PWA) that leverages the underlying microservice-based backend. In our experimental deployment, posters with QR-codes were put up around the university campus for users to scan, after which they are presented with tasks to complete to earn credits. They can then redeem these credits at the coin dispensing machine we installed next to an arcade machine for rewards. We demonstrate a typical user workflow in the QRowdsource application:
\begin{enumerate}

\begin{figure}[hbt!]
\subfloat{\includegraphics[height=6cm]{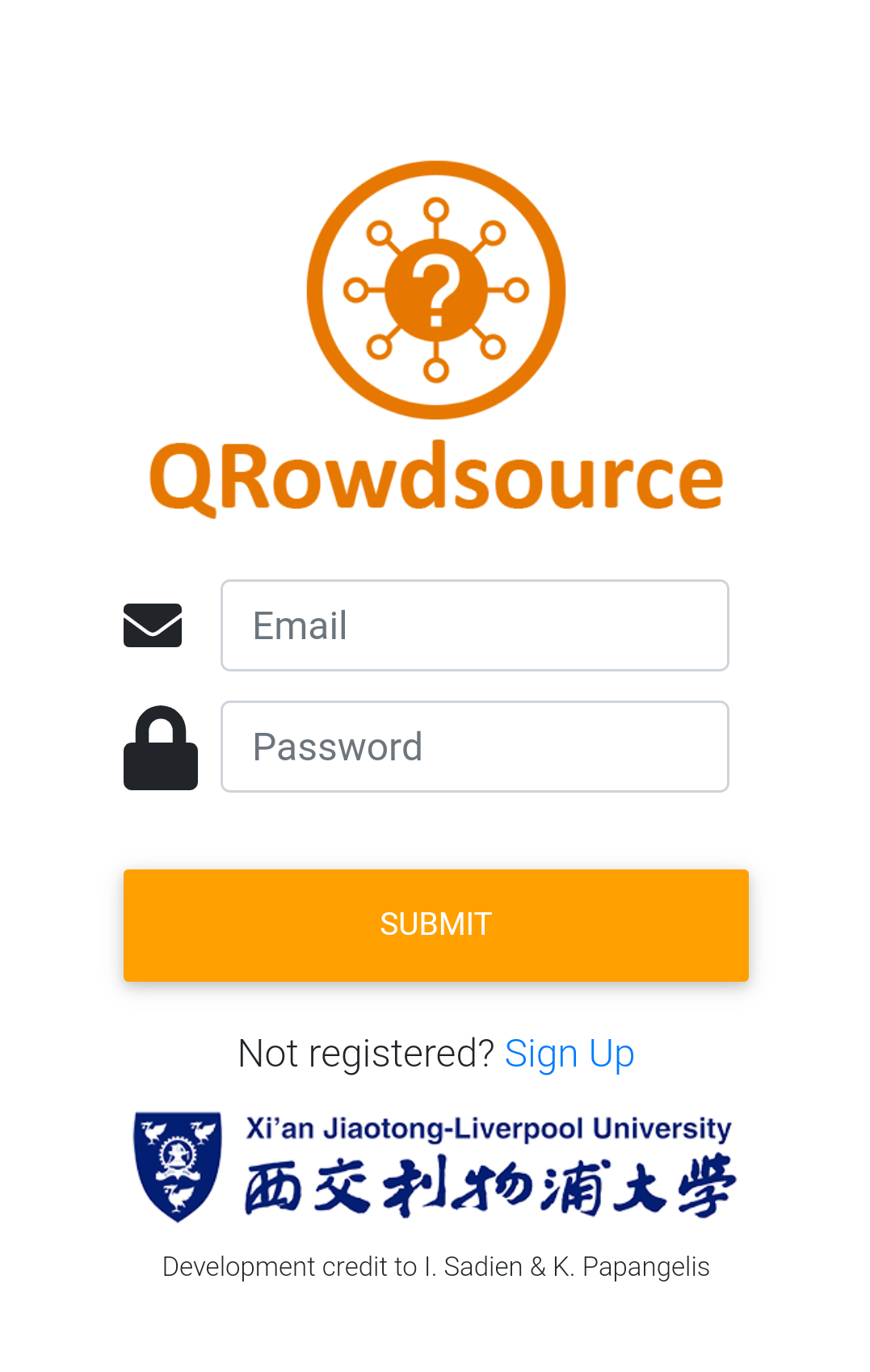}}
\hfill
\subfloat{\includegraphics[height=6cm]{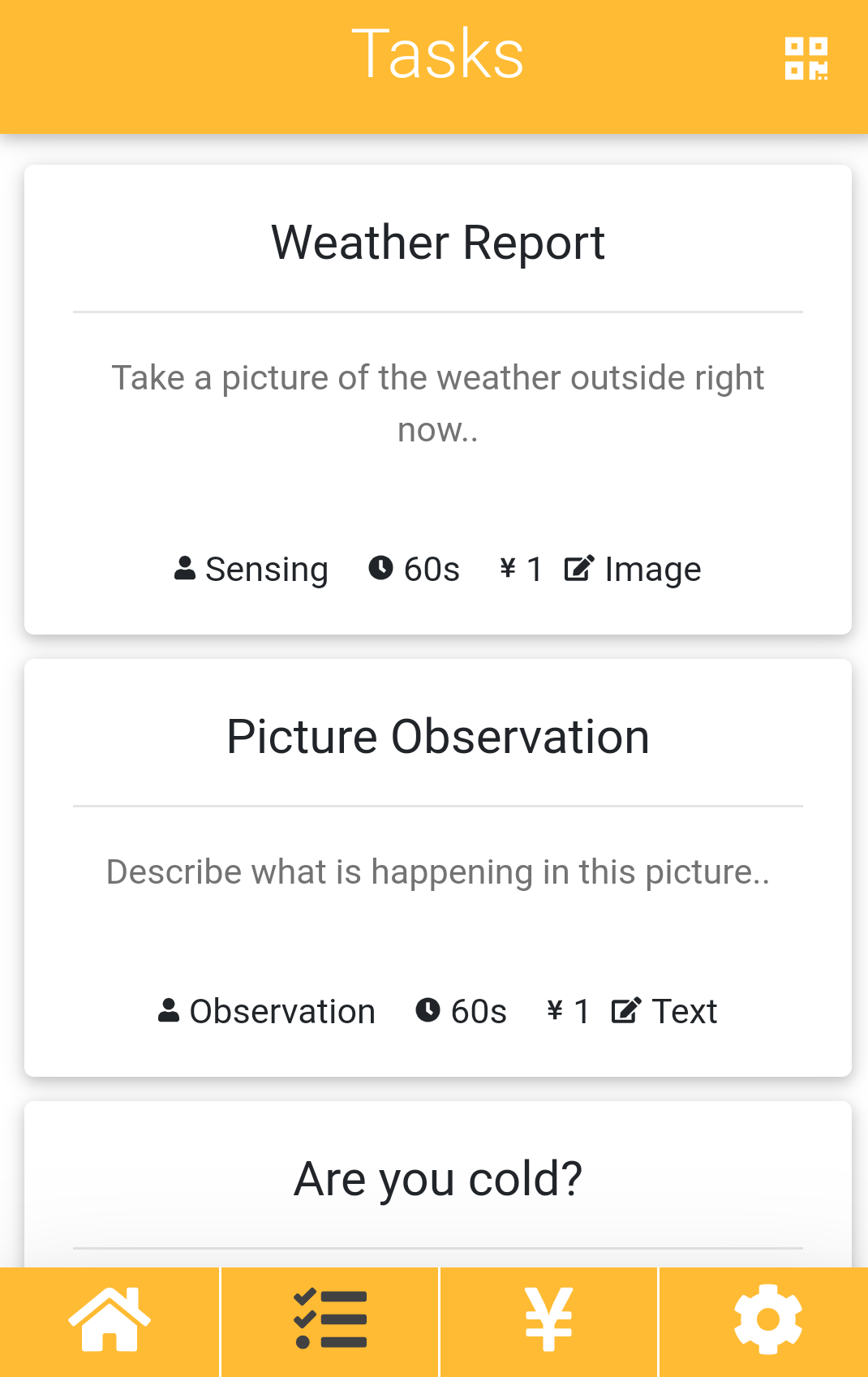}}
\hfill
\caption{Left: User login screen. Right: Task selection.}~\label{fig:userandtasking}
\end{figure}

\item Users start by scanning a QR-code at any of the available locations to access the QRowdsource platform. They then register their email addresses to create a profile within the QRowdsource platform. If they already have an account, users can login in with their email and password to access the platform.
\item Users then navigate to tasking portion of the application, where they are presented with tasks specific to the location at which they scanned the QR-code. Tasks are presented in a scrolling interface with a title, description, and information about their difficulty, response type, reward amount, and date posted.
\begin{figure}[hbt!]
\centering
\subfloat{\includegraphics[height=6cm]{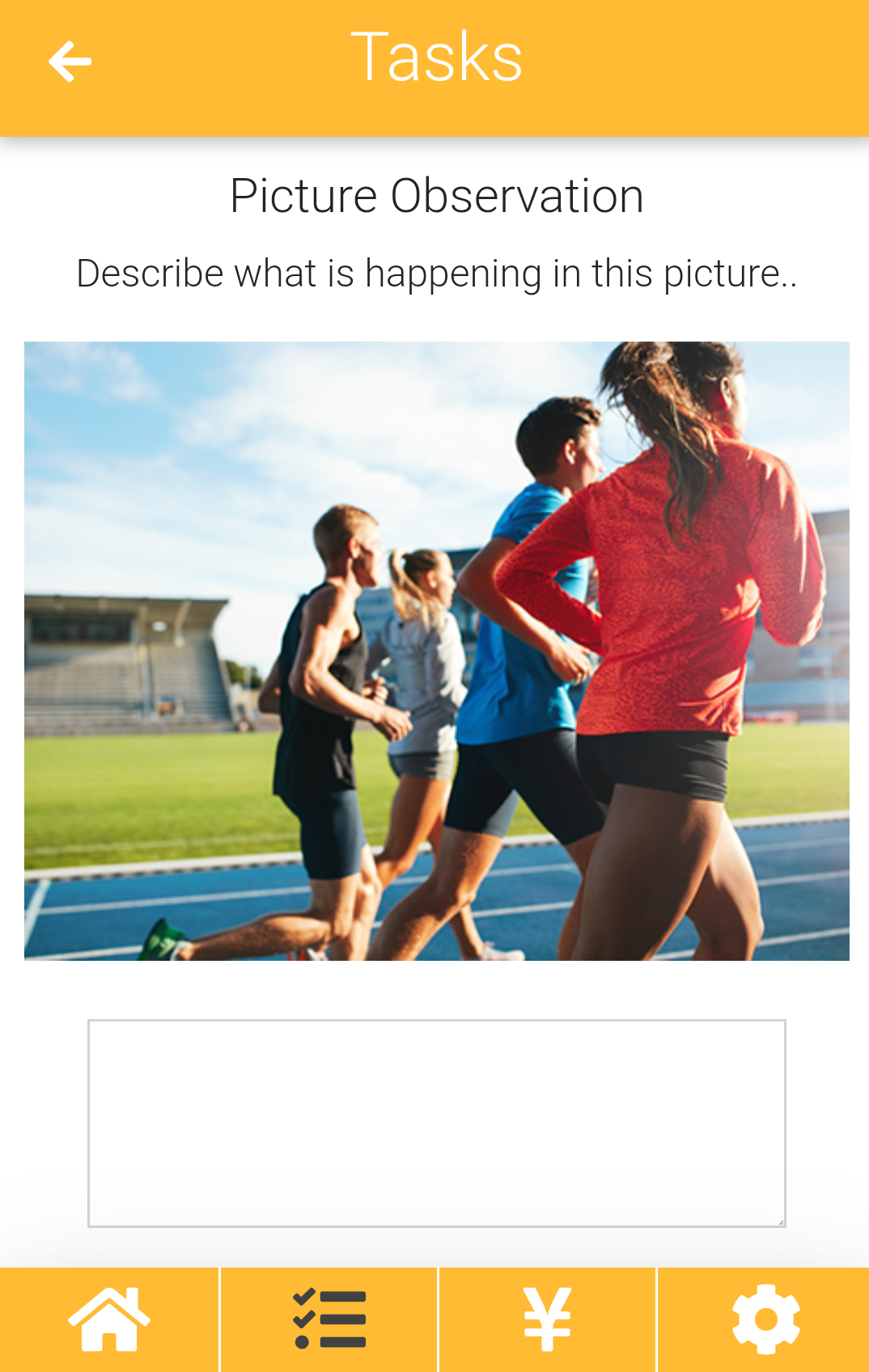}}
\hfill
\subfloat{\includegraphics[height=6cm]{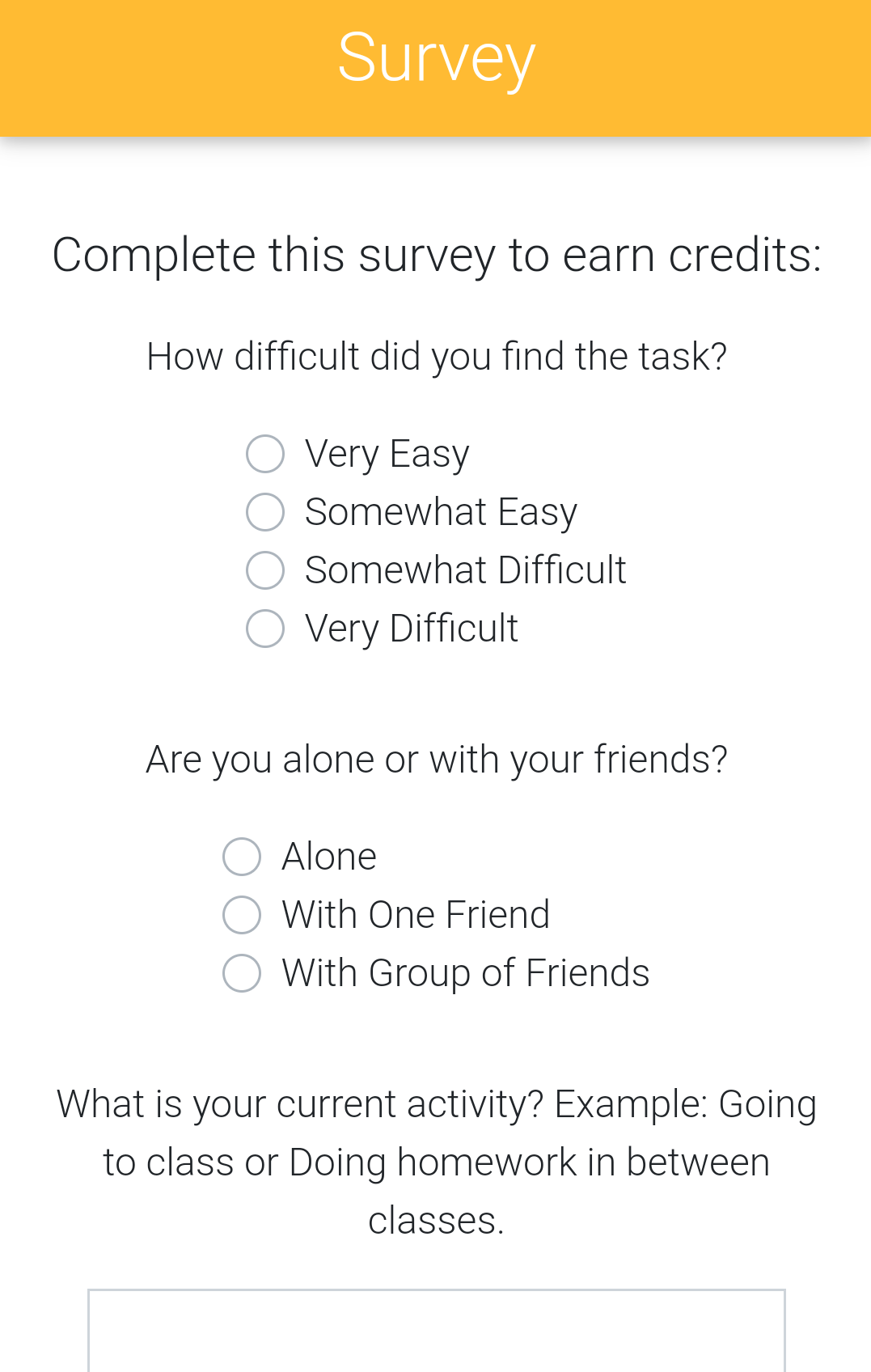}}
\hfill
\caption{Left: Selected task. Right: ESM questionnaire.}~\label{fig:esm}
\end{figure}
\item When a task is selected, the user is presented with input options to complete the task. Metrics pertaining to task performance are recorded as the user interacts with the tasks.
\item Upon task completion, an ESM questionnaire is shown to the user in order to gain information about the user's context at the time of interaction with the platform. Once this is completed, the user is awarded a number of credits depending on the value of the task chosen.

\begin{figure}[hbt!]
\centering
\subfloat{\includegraphics[height=6cm]{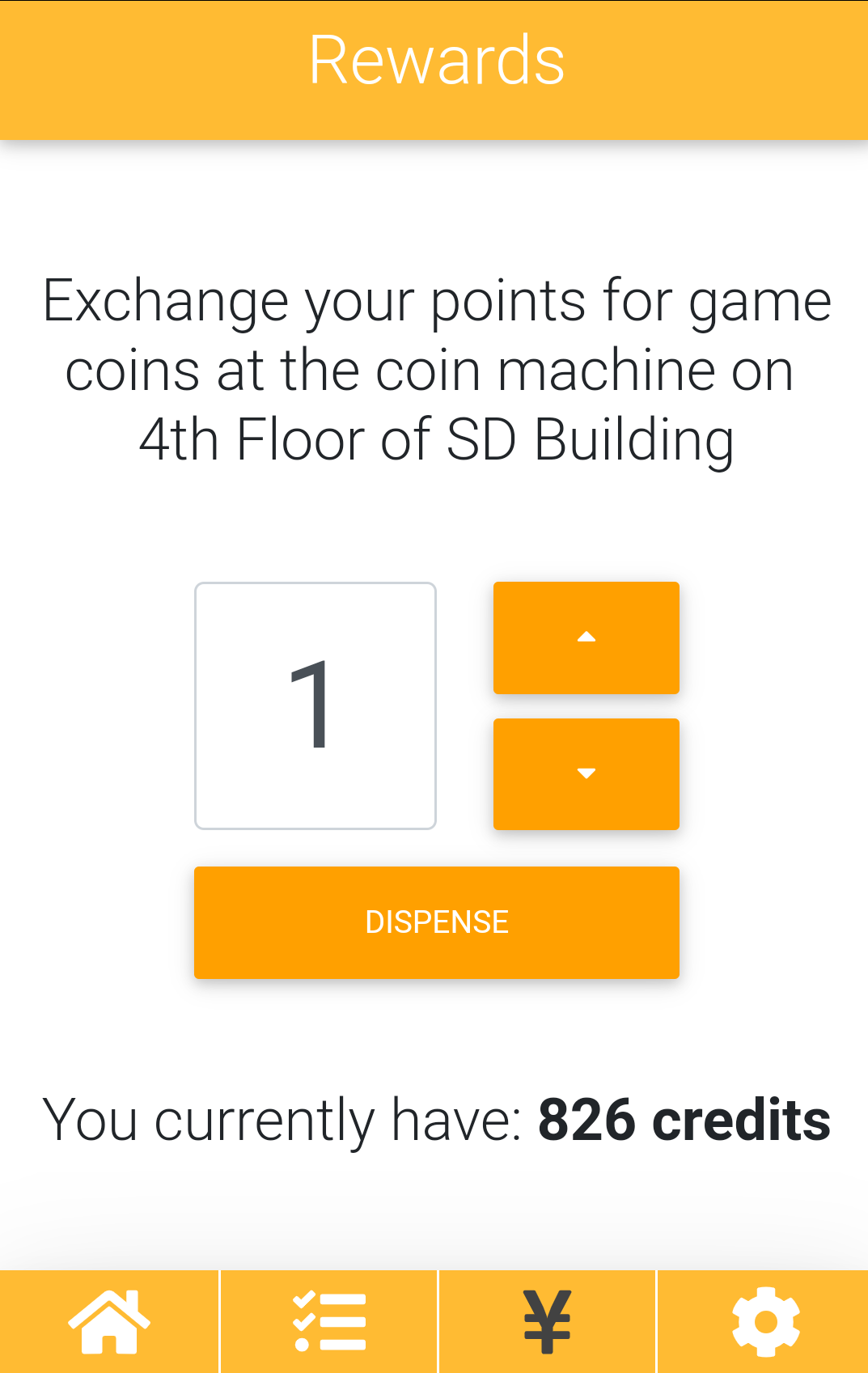}}
\hfill
\subfloat{\includegraphics[height=6cm]{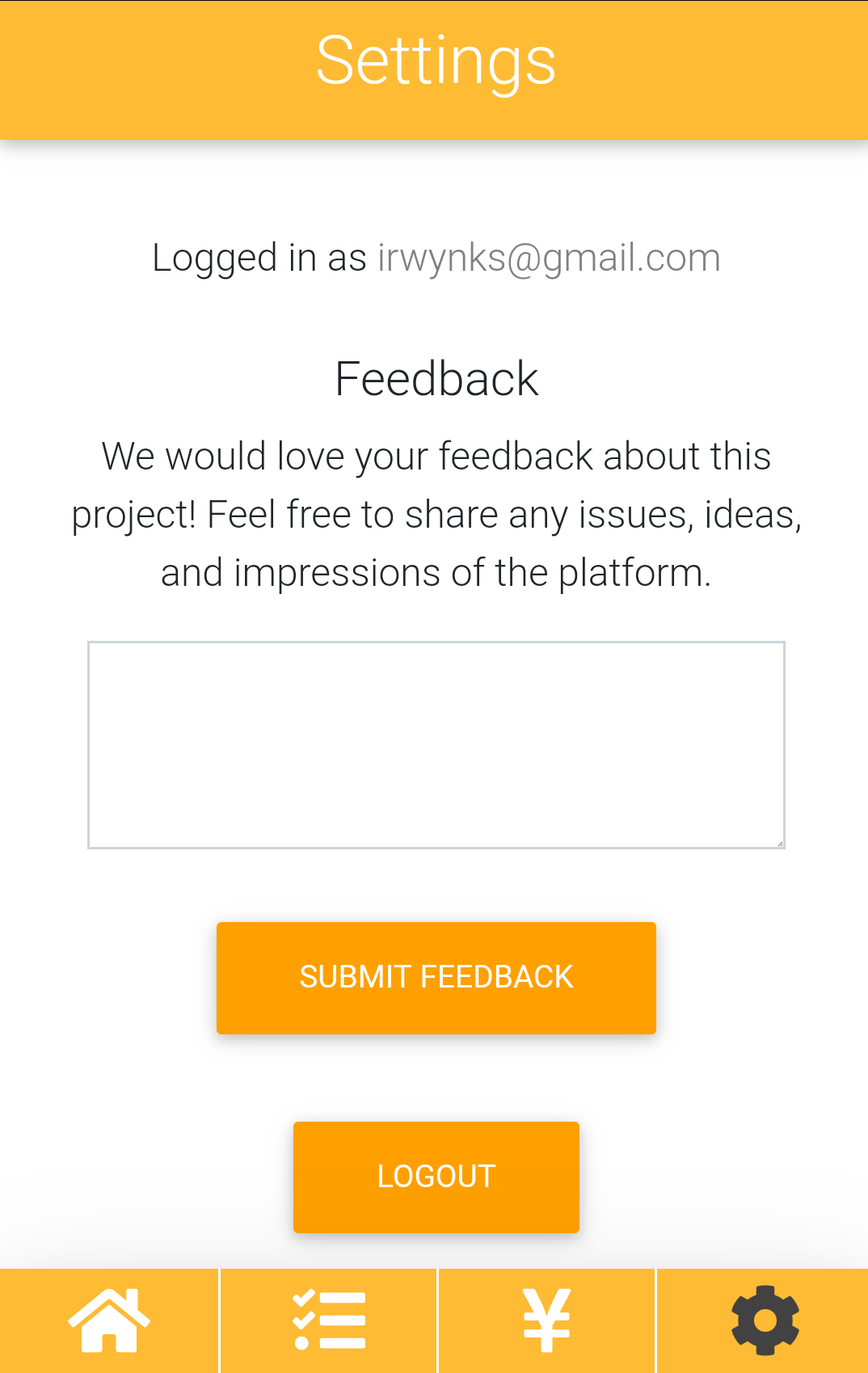}}
\hfill
\caption{Left: Redeem rewards. Right: Feedback and logout.}~\label{fig:rewardandfeedback}
\end{figure}
\item The user can then redeem their earned credits for game coins through the app while being physically present at the coin dispensing machine. These are used to play games at the arcade machine deployed for the QRowdsource experiment.
\item Should they wish to do so, the users can provide feedback about issues or their experience with the platform. They can also logout of the platform once they are done.
\end{enumerate}

\section{Microservice Design and Architecture}
In this section we step through the design process that was undertaken in developing the QRowdsource experimentation platform. This being the first iteration of QRowdsource, we made certain decisions pertaining to our implementation that simplified the deployment process.
\subsubsection{Service Discovery}
We decided to forgo service discovery for our first prototype and instead relied on a single consolidated network environment in which we deployed our services as process instances. While it is an important part of large scale deployments with dynamic network locations, our use case did not require that added level of complexity.

\subsubsection{Containerization and Automation}
Due to the relatively small size of our deployment, we deemed it unnecessary to make use of containers and instead decided to user PM2 as a process manager/load balancer and manually scaled the service instances up and down depending on load.

\subsection{Functional Decomposition}
We first decomposed QRowdsource into a set of essential features in order to determine the functional units to be implemented as core and orchestration services.

\subsubsection{Authentication}
Authentication is vital part of any application accessible over the internet, enforcing both security and resource access rules based on user roles. We chose to use JWT tokens for authentication given the fact that the frontend of the QRowdsource application was built as a PWA for mobile devices, and deployed this service in the edge layer along with the API gateway (Figure~\ref{fig:edgeLayer}).
\subsubsection{User Management}
User management is necessary to facilitate the storage and retrieval of user accounts and associated user information, such as number of tasks completed, last login, and credits earned. This feature also allows users to manage and customize their own profiles.
\subsubsection{Tasking}
Tasking is central to the concept of crowdsourcing. Required functionality includes the management of tasks from an administrative perspective, task allocation depending on location, and data collection for completed tasks.
\subsubsection{User Interactions and Task Performance Metrics}
Specific to QRowdsource is the need to measure task performance metrics such as the time to completion for specific tasks; and the number of times a task has been seen, selected, completed, and/or dropped. Additionally, user interactions with the application need to be logged in order to obtain insight on behavioral patterns.
\subsubsection{ESM and User Feedback}
Once again specific to QRowdsource is the need to obtain user context upon completion of tasks. This takes the form of an ESM questionnaire that is shown when a task is completed. Since QRowdsource is an experimentation platform, it is also important to have a channel of communication between users and researchers, which we implement as a user feedback form.
\subsubsection{Location}
The spatial dimension of the work carried out by users in the QRowdsource platform makes it important to have the appropriate services in place to validate congruence between location data from the users' device sensors and the coordinates that have been assigned to the tasks that they complete.
\subsubsection{Rewarding}
Rewards are the motivating force behind crowdsourcing and can either be extrinsic (e.g. money, digital currency, etc.) or intrinsic (e.g. altruism, social recognition, etc.). In the QRowdsource platform, we use entertainment as incentive: users earn credits to be redeemed at a coin dispensing machine in order to play games. Thus, there is a need to interface with the coin dispensing machine and also keep track of user credits.
\subsubsection{Reporting}
Being able to extract meaningful information from the data is key to any research platform, and thus it must be possible to query the stored data across user interactions, responses, and task performance metrics.

\begin{figure}[h!]
\centering
  \includegraphics[width=1\columnwidth]{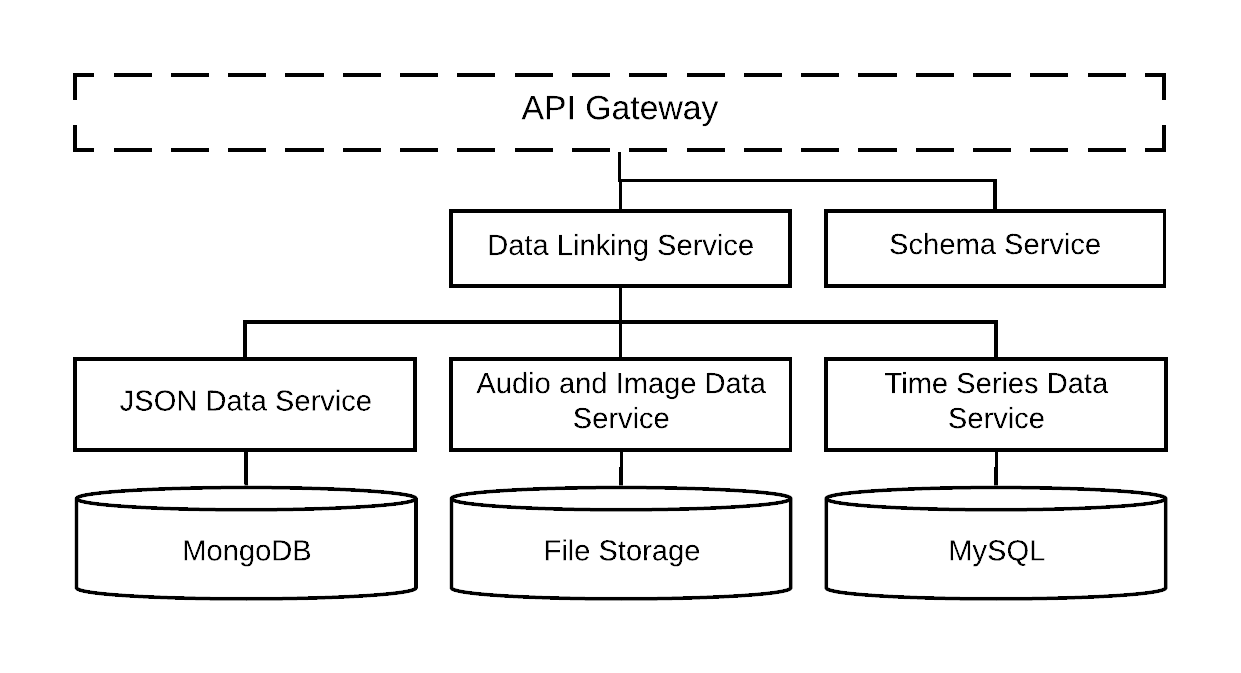}
  \caption{Organization of the data orchestration services in the QRowdsource platform.}~\label{fig:dataLayer}
\end{figure}
\subsection{Data Considerations}
Following from the generic microservices model developed by Braun et al. for environmental data management \cite{Braun2017AManagement}, we make use of the architecture depicted in Figure \ref{fig:dataLayer} for our data orchestration requirements. In our case, metrics for performance aspects of different tasks (task completion time, number of times seen, selected, completed etc.) are stored as time-series data, which is best served by relational databases and implemented using MySQL. Sensor output from mobile devices and textual responses from users vary in size and structure, and lend themselves well to document stores with flexible schemas such as mongoDB, which we implemented separately from the time-series data store. Finally, we built an independent service for image and audio file storage to make it possible to include additional steps such as compression and format conversion. In order to keep track of the different data associations across our databases, we also built a data linkage service to resolve the links embedded in our data. Finally, in order to enforce a bounded context and to ensure that all the services have access to the same database schemas, a schema sharing service with service level access control was created.

\subsection{Software Stack}
Our software stack was hosted on a single Digital Ocean droplet (VPS instance) and served behind a nginx web server configured as a reverse proxy with HTTPS enabled for security. To facilitate development and maintenance, we focused on using JavaScript technologies throughout the whole software stack. The microservices and gateway were implemented using node.js, a lightweight, single threaded open source server-side environment; and the frontends for both the QRowdsource PWA and administration interface were built using react.js, a popular JavaScript library used for creating dynamic, stateful user interfaces. MongoDB was used as database for persistent object storage for the scalability and versatility of its schema-less document storage format, while MySQL was used for fast time-series data storage made possible by its relational model. Communication between services and external applications was implemented using a combination of REST and WebSockets/RabbitMQ for synchronous and asynchronous messaging respectively. Finally, we used PM2, a process manager for node.js, to take care of instance spawning and load balancing.

\subsection{Tying Everything Together}
\begin{figure}[hbt!]
\centering
  \includegraphics[height=7cm, angle=-90]{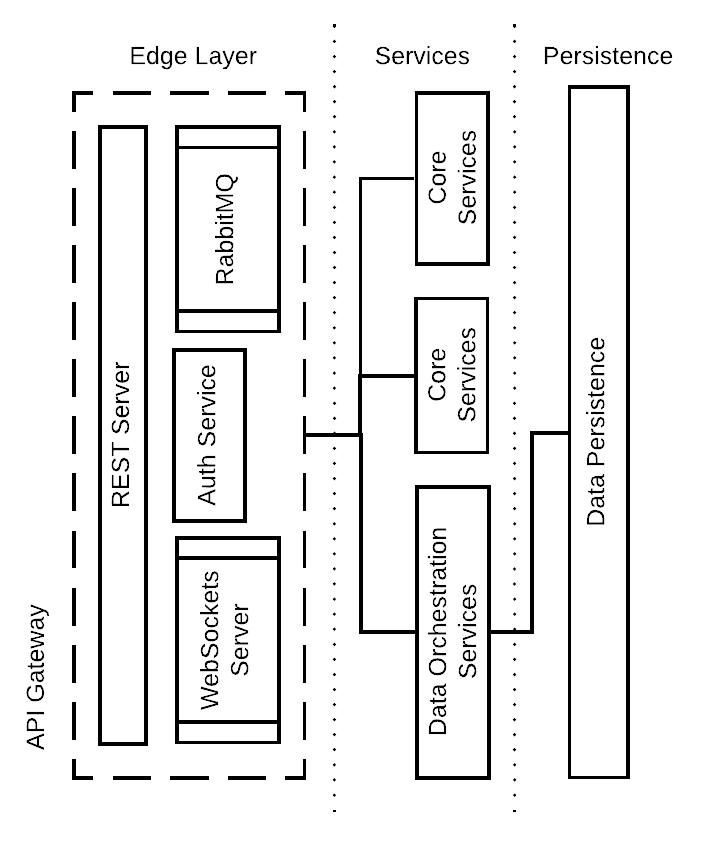}
  \caption{API gateway deployed in the edge layer to handle communication.}~\label{fig:edgeLayer}
\end{figure}

The key components in any microservice architecture are the communication providers that take are of messaging between services and requests and responses to and from the platform itself. This functionality is managed by the API gateway that we deploy in the edge layer (Figure~\ref{fig:edgeLayer}) and implemented using REST (internal and external synchronous communication), WebSockets (external asynchronous communication), and RabbitMQ (internal asynchronous communication).

\section{The good and the bad}
Software development using a microservices pattern presents a number of advantages and disadvantages when compared to a monolithic architecture. We discuss these in the context of developing an experimentation framework for MLBC.

\subsection{Advantages}
\subsubsection{Modularity}
Modularity is one of the salient points of the QRowdsource experimentation platform; the possibilities for adapting QRowdsource to different research experiments are endless. Numerous changes were made during the development process that, had we been using a monolithic architecture, would have required a complete refactoring of the code base. With the microservice architecture, adding new features or modifying existing functionality is as simple as routing or rerouting messages and requests to a different module.
\subsubsection{Minimal Downtime}
As an experimentation platform, the inner workings of QRowdsource are subject to change depending on the requirements (and whim) of those running the research experiments. With a microservice based approach, it was possible to achieve close to zero downtime throughout the duration of our live deployment, even when making changes to the core functionality of the platform while users were actively performing crowdsourcing tasks.
\subsubsection{Distributed Development}
The modular nature of microservice architecture also made is possible to distribute development work to teams with skillsets that matched the services being worked on. The fact that communication between different services is standardized meant that different teams could implement their application and data logic however they saw fit. This also made the QRowdsource platform easier to debug and maintain across the board. Most importantly, this means that any interested party can contribute to the development of the QRowdsource platform.
\subsubsection{Variance}
The microservice approach made it possible to use the most appropriate technologies in order to achieve the desired functionality. We expect different deployments and configurations to require their own set of specific modules and implementations. This versatility makes it possible to pick and choose the right tools for the job instead of having to stick to a single software stack, both at inception and over time as experiment requirements evolve.
\subsubsection{Data Transparency and Availability}
One of the great advantages of using a microservice architecture is that data can be transparently stored to and retrieved from a number of sources and be made accessible through a standard interface. Microservice architectures are nowadays being used in all kind of software projects, from games \cite{papangelis_conquering_2017, papangelis_get_2017, 10.1007/978-3-030-14745-7_16} to social-networks \cite{papangelis_unfolding_2017} and everything in between \cite{chamberlain_sounding_2018, chamberlain_mapping_2017}. Deploying reporting dashboards and data extraction tools is a breeze, and this greatly facilitated the statistical analysis of the data collected through the experimentation platform during our deployment.
\subsubsection{Scalability}
As traffic increased with user adoption during the deployment of our first research experiment, it was a simple matter of spawning additional instances of the services experiencing heavy load and balancing the network requests going to them to mitigate network slowdown. This is something that would have required the duplication of the entire platform, along with the additional storage, CPU, and memory overheads, had QRowdsource been implemented using a monolithic architecture.

\subsection{Disadvantages}
\subsubsection{Orchestration}
One of the major challenges we encountered in building QRowdsource was the orchestration of services in implementing the desired functionalities. The distributed nature of the microservice pattern adds extra layers of complexity in organizing the communication between services. This may prove to be a barrier for entry as researchers will need to learn how to connect/reorganize existing and custom built modules in the experimentation framework.
\subsubsection{Communication Overhead}
As we move away from a monolithic architecture, additional overhead is created by the messaging systems needed to enable communication between the different services in the application framework. As the QRowdsource ecosystem grows and additional modules and services are created, this may prove to be an issue.
\subsubsection{Variance}
With greater freedom of choice and diversity of technologies comes the added burden of having to manage them independently, which means additional software and tooling to implement. A lot of time was spent learning about and implementing the supporting technologies involved in deploying a microservice based platform. In order to alleviate this to some extent we opted to develop QRowdsource mostly around JavaScript and tried to keep diversity to a minimum. However, the increase in variance is inevitable over time as the platform grows and additional features are added.

\section{Lessons learned in developing QRowdsource}
In this section we share a number of lessons learned in building an MLBC experimentation platform using a microservice architecture:

\subsubsection{Services don't have to be singular in purpose}
While there is an ongoing debate about the size of a microservice \cite{LewisJamesFowler2014MicroservicesTerm}\cite{Newman2015BuildingMicroservices}, the the general consensus is that services should be stateless and serve a single function. However, during the design of the QRowdsource platform, we found ourselves merging multiple services together in order to reduce the complexity of the framework and the number of interconnects required. This was done in a domain driven manner in order to retain context boundedness in our services; as such, while maybe not singular in purpose, the services implemented in the core of the QRowdsource framework stay true to the microservices pattern.

\subsubsection{Shared Data Layer: A microservice anti-pattern}
The primary goals in moving from a monolithic approach to a microservice architecture are  1) to split the functionality into small, single-purpose services, and 2) to divide the monolithic data store into a number of small databases owned by these services. While it is a logical approach, this level of separation raises a number of problems when it comes to data migrations and future integrations \cite{Richards2016MicroservicesPitfalls}. In designing the QRowdsource platform, we decided to prioritize expandability, and thus separated the data persistence layer into its own set of services for data management and schema sharing. Although essentially an anti-pattern, this approach made it easier to develop and implement modules on top of the core framework as database operations and access control mechanisms are normalized across the platform. The only drawback is that changes and additions to schemas automatically introduce a certain amount of downtime, though once again thanks to the microservice approach, this offline time (for the schema sharing service) is kept to a minimum.

\section{Conclusion and Future Work}
The need for standardized and evolutionary experimentation tools and platforms is undeniable in the pursuit of further research in MLBC. In developing QRowdsource, and more importantly in designing the underlying microservices framework upon which QRowdsource is based, we attempt to provide a starting point for the creation of such tools. In this paper, we provide a functional decomposition of the main processes in MLBC and break these down into components to be implemented as services within a microservice architecture. We then take a step back to talk about the supporting infrastructure and how the services are interconnected and deployed, giving an example of technologies that could be used in doing so using our specific deployment. Finally, we take a look at how the different advantages and disadvantages of microservices architectures translate in the context of an experimentation framework, and in our lessons learned section, we further explore some of the design decisions made in the process of developing QRowdsource, reflecting upon how these fit within the paradigm of microservices.

Future work involves further developing  QRowdsource, as well creating additional functionality. This for example, may include containerization and service discovery. Further to this, we will need to test QRowdsource in a "real world" setting to ensure its reliability before releasing a beta version of it for upcoming experiments.

\bibliographystyle{SIGCHI-Reference-Format}
\bibliography{refbib}

\end{document}